\documentstyle[11pt,aasms4,flushrt,tighten]{article}
\singlespace

\def\kmsmpc{\,{\rm km\,s^{-1}\,Mpc^{-1}}}

\def\msun{\,{\rm M_\odot}}

\def\sfrd{\,{\rm M_\odot\,yr^{-1}\,Mpc^{-3}}}
\def\rd{\,{\rm yr^{-1}\,Mpc^{-3}}}

\def\etal{{et al.\ }}

\def\spose#1{\hbox to 0pt{#1\hss}}
\def\lta{\mathrel{\spose{\lower 3pt\hbox{$\mathchar"218$}}
     \raise 2.0pt\hbox{$\mathchar"13C$}}}
\def\gta{\mathrel{\spose{\lower 3pt\hbox{$\mathchar"218$}}
     \raise 2.0pt\hbox{$\mathchar"13E$}}}
 
\begin{document}
\title{ON THE EVOLUTION OF THE COSMIC SUPERNOVA RATES}

\author{Piero Madau}

\affil{Space Telescope Science Institute, 3700 San Martin Drive,
Baltimore MD 21218;\ madau@stsci.edu}

\author{Massimo Della Valle}

\affil{Dipartimento di Astronomia, Universit\`a di Padova, Vicolo dell'
Osservatorio 5, I-35122 Padua, Italy;\ dellavalle@astrpd.pd.astro.it} 

\and

\author{Nino Panagia\altaffilmark{1}}

\affil{Space Telescope Science Institute, 3700 San Martin Drive,
Baltimore MD 21218;\ panagia@stsci.edu}

\altaffiltext{1}{Affiliated with the Astrophysics Division, Space Science 
Department of ESA}

\begin{abstract}
Ongoing searches for supernovae (SNe) at cosmological distances have recently
started to provide a link between SN Ia statistics and galaxy evolution. We use
recent estimates of the global history of star formation to compute the
theoretical Type Ia and Type II SN rates as a function of cosmic time from the
present epoch to high redshifts. We show that accurate measurements of
the frequency of SN events in the range $0<z<1$ will be valuable probes of the
nature of Type Ia progenitors and the evolution of the stellar birthrate in the
universe. The {\it Next Generation Space Telescope} should detect of order 20 
Type II SNe per $4'\times 4'$ field per year in the interval $1<z<4$. 
\end{abstract}


\section{Introduction}

The remarkable progress in our understanding of faint galaxy data made possible
by the combination of HST deep imaging (Williams \etal 1996) and 
ground-based spectroscopy (Lilly \etal 1995; Ellis \etal 1996; Cowie \etal
1996; Steidel \etal 1996), has recently permitted to shed
some light on the evolution of the stellar birthrate in the universe, to 
identify the epoch $1\lta z\lta 2$ where most of the optical extragalactic 
background light was produced, and to set important contraints on galaxy 
formation scenarios (Madau \etal 1998; Steidel \etal 1998). While one of the
biggest uncertainties in our knowledge of the emission history of the universe
is probably represented by the poorly constrained amount of starlight that 
was absorbed by dust and reradiated in the IR at early and late epochs,
one could also imagine the existence of a large population of relatively
old or faint galaxies still undetected at high-$z$, as the color-selected 
ground-based and {\it Hubble Deep Field} samples include only the most 
actively star-forming young objects.
It is then important at this stage to devise different  
observational strategies, free of some of the biases that plague current galaxy
surveys, and to make testable predictions for future astronomical 
capabilities, such as SIRTF, FIRST and NGST.

Here, we shall focus our attention on the rate of supernova (SN) 
explosions in the universe. An obvious reason to consider SNe is purely 
observational, i.e. the fact that they are very bright objects, with 
luminosities as high as $10^{10}~L_\odot$, and are point-like
sources, making their detection possible even at very large 
distances/redshifts.  More in general, the evolution of the SN rate 
with redshift contains unique
information on the star formation history of the universe, the initial mass
function (IMF) of stars, and the nature of the binary companion in Type Ia
events. All are essential ingredients for understanding galaxy formation,
cosmic chemical evolution, and the mechanisms which determined the efficiency
of the conversion of gas into stars in galaxies at various epochs (e.g. Madau
\etal 1996; Madau, Pozzetti, \& Dickinson 1997; Renzini 1997). While the
frequency of ``core-collapse supernovae'', SN~II and possibly SN~Ib/c, which
have short-lived progenitors (e.g. Wheeler \& Swartz 1993) is
essentially related, for a given IMF, to the instantaneous stellar birthrate of
massive stars, Type Ia SNe -- which are believed to result from the
thermonuclear disruption of C-O white dwarfs in binary systems -- follow a
slower evolutionary clock, and can then be used as a probe of the past history
of star formation in galaxies (e.g. Branch \etal 1995; Ruiz-Lapuente, Canal, \&
Burkert 1997; Yungelson \& Livio 1998). 
The recent detection of Type Ia SNe at cosmological distances (Kim \etal 
1997; Garnavich \etal 1998; Perlmutter \etal 1998) allow for the first time a detailed comparison between the SN rates
self-consistently predicted by stellar evolution models that reproduce the
optical spectrophotometric properties of field galaxies,
and the observed values. 

In this {\it Letter} we show how accurate measurements
at low and intermediate redshifts  of the frequencies
of Type II(+Ib/c) and Ia SNe could be used as an independent test for the star
formation and heavy element enrichment history of the universe,  and 
significantly improve our understanding of the intrinsic nature and age 
of the populations involved in the SN explosions. A determination of the amount 
of star formation at early epochs is of crucial importance, as the two 
competing scenarios for galaxy formation, monolithic collapse -- where 
spheroidal systems formed early and rapidly, experiencing a bright starburst 
phase at high-$z$ (Eggen, Lynden-Bell, \& Sandage 1962; Tinsley \& Gunn 
1976) -- and hierarchical clustering -- where ellipticals form continuosly by 
the merger of disk/bulge systems (White \& Frenk 1991; Kauffmann \etal 1993) 
and most galaxies never
experience star formation rates in excess of a few solar masses per year (Baugh
\etal 1998) -- appear to make rather different predictions in this regard. We show how,
by detecting Type II SNe at high-$z$, the {\it Next Generation Space Telescope}
should provide an important test for distinguishing between different 
scenarios of galaxy formation.

\section{Basic Theory}

\subsection{Cosmic Star Formation History}

In this section we shall follow Madau \etal (1998), and model the emission
history of field galaxies at ultraviolet, optical, and near-infrared
wavelengths by tracing the evolution with cosmic time of their luminosity
density, 
\begin{equation}
\rho_\nu(z)=\int_0^\infty dL_\nu L_\nu y(L_\nu,z)=\Gamma(2+\alpha)y_*L_*,
\end{equation}
where $y(L_\nu,z)$ is the best-fit Schechter luminosity function in each
redshift bin. The integrated light radiated per unit volume from the entire
galaxy population is an average over cosmic time of the stochastic, possibly
short-lived star formation episodes of individual galaxies, and follows a
relatively simple dependence on redshift. Madau \etal (1998) have shown how 
a stellar evolution model, defined by a time-dependent star formation 
rate per unit volume, $\psi(t)$, a universal IMF, $\phi(m)$, and some amount 
of reddening, can actually reproduce the optical data reasonably well. In such a
system, the luminosity density at time $t$ is given by the convolution integral
\begin{equation} 
\rho_\nu(t)=p_{\rm esc}\int^t_0 l_\nu(t')\psi(t-t')dt', 
\end{equation} 
where $l_\nu(t')$ is the specific luminosity radiated per unit initial mass
by a generation of stars with age $t'$, and $p_{\rm esc}$ is a
time-independent term equal to the fraction of emitted photons which are not
absorbed by dust, $p_{\rm esc}=\exp(-\tau_\nu)$ in a foreground screen model.
The function $\psi(t)$ is derived from the observed UV luminosity density, and
is then used as input to the population synthesis code of Bruzual \& Charlot 
(1998). 

Figure 1{\it a} shows the model predictions for the evolution of $\rho_\nu$ for
a Salpeter (1955) function ($\phi\propto m^{-2.35}$), $E_{\rm B-V}=0.1$ 
with SMC-type dust (in this case, the
observed UV luminosities must be corrected upwards by a factor of 1.4 at 2800
\AA\, and 2.1 at 1500 \AA), and a star formation history which traces the rise,
peak, and sharp drop of the UV emissivity. \footnote{A convenient fit to 5\% 
accuracy to the star formation history used for Figure 1{\it a}
is $\psi(t)=0.049[t_9^5 e^{-t_9/0.64}+0.2(1-e^{-t_9/0.64})] \sfrd$, where 
$t_9$ is the Hubble time in Gyr, $t_9=13(1+z)^{-3/2}$. Note that, although in 
our calculations
the IMF extends down to $m_l=0.1 \msun$, stars below 0.8 $\msun$ make only a
small contribution to the emitted near-IR light. This introduces a
non-negligible uncertainty in our estimates of the total stellar birthrate. For
example, in the case of a Salpeter IMF with $m_l=0.5 \msun$, the inferred star
formation rate, $\psi\propto (m_l^{-0.35}-m_u^{-0.35})$, would decrease 
by a factor of 1.88.}~ For simplicity, the metallicity was fixed to 
solar values and the IMF
truncated at 0.1 and 125 $\msun$: none of these assumptions is crucial for our
results. The data points show the observed luminosity
density in six broad passbands centered around 0.15, 0.20, 0.28, 0.44, 1.0, and
2.2 \micron. The model is able to account for the entire background light
recorded in the galaxy counts down to the very faint magnitude levels probed by
the {\it Hubble Deep Field} (HDF), and produces visible mass-to-light ratios at
the present epoch which are consistent with the values observed in nearby
galaxies of various morphological types. The bulk ($\gta 60\%$ by mass) of the stars present today formed
relatively recently ($z\lta 1.5$), consistently with the expectations from a
broad class of hierarchical clustering cosmologies (Baugh \etal 1998), and in
good agreement with the low level of metal enrichment observed at high
redshifts in the damped Lyman-$\alpha$ systems (Pettini \etal 1997). 

One of the biggest uncertainties in our understanding of the evolution of
luminous matter in the universe is represented by the poorly constrained
amount of starlight that was absorbed by dust and reradiated in the far-IR
at early epochs. While the Salpeter IMF, $E_{\rm B-V}=0.1$ SMC-dust model 
reproduces
quite well the rest-frame UV colors of high-$z$ objects in the HDF, the 
prescription for a ``correct'' de-reddening of Lyman-break galaxies 
is the subject of an ongoing debate (see Pettini \etal 1998 and references
therein). Figure 1{\it b} shows the model predictions for a monolithic collapse 
scenario, where half of the present-day stars -- the fraction contained in
spheroidal systems (Schechter \& Dressler 1987) -- were formed at $z>2.5$ and
were enshrouded by dust.\footnote{A fitting formula to the stellar birthrate 
used for Figure 1{\it b} is $\psi(t)=0.336e^{-t_9/1.6}+0.0074(1-e^{-t_9/
0.64})+0.0197t_9^5e^{-t_9/0.64} \sfrd$ to 1\% accuracy.}~ Consistency with 
the HDF ``dropout'' analysis has been obtained
assuming a dust extinction which increases rapidly with redshift, 
$E_{\rm B-V}=0.011(1+z)^{2.2}$. This results in a correction to the rate of 
star formation by a factor of $\approx 2, 5,$ and 15 at $z=2, 3$ and 4, 
respectively. 
The model is still consistent with the global history of light,
but overpredicts the metal mass density at high redshifts as sampled by QSO
absorbers. In order for such a model to be acceptable, the gas traced by the
damped Lyman-$\alpha$ systems would have to be physically distinct from the 
luminous star forming regions observed in high-$z$ galaxies, and to be
substantially under-enriched in metals compared to the cosmic mean.

In the next section we shall compute the expected evolution with cosmic time of
the Type Ia and II/Ib,c supernova frequencies for the two star formation
histories -- for brevity, we shall refer to them as the ``merging'' versus
``monolithic collapse'' scenarios -- discussed above. By focusing on the 
integrated
light radiated by the galaxy population as a whole, our approach will not
specifically address the evolution and the SN rates of particular subclasses of
objects, like the oldest ellipticals or low-surface brightness galaxies, whose
star formation history may have differed significantly from the global average.

\subsection{Type Ia and II(+Ib/c) Supernova Rates}

Single stars with mass $>8\msun$ evolve rapidly ($\lta 50\,$ Myr) through all
phases of central nuclear burning, ending their life as Type II SNe with
different characteristics depending on the progenitor mass (e.g. Iben \&
Renzini 1983). For a Salpeter IMF (with lower and upper mass cutoffs of 0.1 
and 125 $\msun$), the core-collapse supernova rate can be related to the 
stellar birthrate according to 
\begin{equation}
{\rm SNR}_{\rm II}(t)=\psi(t){\int_8^{125} dm \phi(m)\over \int_{0.1}^{125} 
dm m \phi(m)}=0.0074\times \left[{\psi(t)\over \sfrd} \right]\, \rd.
\end{equation} 
It is worth noting at this stage that our model predictions for the frequency
of Type II events are largely independent of the assumed IMF. This follows from
the fact that the rest-frame UV continuum emission -- which is used as an
indicator of the instantaneous star formation rate $\psi(t)$ -- from all but
the oldest galaxies is entirely dominated by massive stars on the main
sequence, the same stars which later give origin to a core collapse SN. 

The specific evolutionary history leading to a Type Ia event remains instead an
unsettled question. SN Ia are believed to result from the explosion of C-O
white dwarfs (WDs) possibly triggered by the accretion of material from a
companion, the nature of which is still unknown (see Ruiz-Lapuente \etal 1997
for a recent review). In a {\it double degenerate} (DD) system, for example,
such elusive companion is another WD: the exploding WD reaches the
Chandrasekhar limit and carbon ignition occurs at its center (Iben \& Tutukov
1984). In the {\it single degenerate} (SD) model instead, the companion is a
nondegenerate, evolved star that fills its Roche lobe and pours hydrogen or
helium onto the WD (Whelan \& Iben 1973; Iben \& Tutukov 1984). While in the
latter the clock for the explosion is set by the lifetime of the primary star
and, e.g., by how long it takes to the companion to evolve and fill its Roche
lobe, in the former it is controlled by the lifetime of the primary star and
by the time it takes to shorten the
separation of the two WDs as a result of gravitational wave radiation. The
evolution of the rate depends then, among other things, on the unknown mass
distribution of the secondary binary components in the SD model, or on the
distribution of the initial separations of the two WDs in the DD model. 

To shed light into the identification issue and, in particular, on the 
clock-mechanism for the explosion of Type Ia's, we shall adopt here a more
empirical approach. We choose to parametrize the rate of Type Ia's in terms of 
a characteristic explosion timescale, $\tau$ -- which defines an explosion
probability per WD assumed to be {\it independent} of time -- and an explosion
efficiency, $\eta$. The former accounts for the time it takes in the various
models to go from a {\it newly born} (primary) WD to the SN explosion itself: a
spread of ``delay'' times results from the combination of a variety of initial
conditions, such as the mass ratio of the binary system, the distribution of
initial separations, the influence of metallicity on the mass transfer rate and
accretion efficiency, etc., folded with the various possible mechanisms that
lead to a Type Ia event. The latter simply accounts for the fraction of stars 
in binary systems that, because of unfavorable
initial conditions, will never undergo a SN Ia explosion. We consider as
possible progenitors all systems in which the primary star has an {\it initial}
mass higher than $m_{\rm min}=3\,M_\odot$ (final mass $\ge 0.72\,M_\odot$,
Weidemann 1987) and lower than $m_{\rm max}=8\,M_\odot$ (cf Greggio \& Renzini
1983): stars less massive
than $3\,M_\odot$ will not produce a catastrophic event even if the companion
has comparable mass, while stars more massive than $8\,M_\odot$ will undergo 
core collapse, generating a Type II explosion. 

With these assumptions the rate of Type Ia events at any one time will be
given by the sum of the explosions of all the binary WDs produced in the
past that have not had the time to explode yet, i.e. 
\begin{equation}
{\rm SNR}_{\rm Ia}(t)={\eta \int^t_0 \psi(t')dt'\int_{m_c} 
^{m_{\rm max}} \exp(-{t-t'-t_m\over \tau})\phi(m)dm 
\over \tau \int\phi(m)dm},
\end{equation}
where $m_c\equiv{\rm max}[m_{\rm min},m(t')]$, $m(t')=(10\,{\rm Gyr}/t')^{0.4}$ 
is the minimum mass of a star that 
reaches the WD phase at time $t'$, and $t_m=10\,{\rm Gyr}/m^{2.5}$ is the 
standard lifetime of a star of mass $m$ (all stellar masses are expressed 
in solar units). For a fixed initial mass $m$, the frequency of Type Ia events
peaks at an epoch that reflect an ``effective'' delay $\Delta t=\tau+t_m$
from stellar birth. A prompter (smaller $\tau$) explosion results in a 
higher SN Ia rate at early epochs. 

\section{Comparison with the Data}

Observed rates of SNe are normally given in units of SNu, one SNu corresponding
to 1 SN per 100 years per $10^{10} L_{B\odot}$. Since a galaxy luminosity
depends on $H_0$, there is a factor $h_{50}^2$ involved in the inferred SN
frequencies. The typical uncertainties on the present-day rates are summarized
in Table 1, where we have reported the determinations of Cappellaro \etal
(1997), Tammann, L\"offler, \& Schr\"oder (1994), and Evans, van den Bergh, \&
McClure (1989). These agree to within a factor of 3 for Type Ia in early type
galaxies and Type Ib/c in spirals, and to within a factor of 2 for Type II in
late spirals. The ``internal'' error in each entry is of order $40\%$. When
weighted according to the local blue luminosity function by spectral type of
Heyl \etal (1997) (which assigns about 28\% of the total local blue luminosity
density to elliptical galaxies, 32\% to early spirals, and 40\% to late
spirals), the tabulated frequencies yield a mean -- averaged over the entire
galaxy population -- SN~Ia rate of (0.12,0.19,0.12) $h_{50}^2$ SNu, and a 
SN~II(+Ib/c) rate of (0.32,0.62,0.60) $h_{50}^2$ SNu (the three values
corresponding to the measurements of Cappellaro \etal 1997, Tammann \etal 1994,
and Evans \etal 1989, respectively). A simple geometric mean of these
determinations gives the values, ${\rm SNR}_{\rm Ia}=0.14\pm 0.06\,h_{50}^2$
SNu and ${\rm SNR}_{\rm II}=0.49\pm 0.2\,h_{50}^2$ SNu, we shall adopt for
comparison with the theoretical rates. 
At higher redshifts, searches for SNe have been pioneered by Hansen \etal
(1989) and Norgaard-Nielson \etal (1989).  Recently, systematic searches of
distant SNe (Kim \etal 1997; Garnavich \etal 1998; Perlmutter \etal 1998) have
provided the first measurement of the rate of Type Ia at $z=0.4$, ${\rm
SNR}_{\rm Ia}=0.21^{+0.17}_{-0.13}\,h_{50}^2$ SNu (Pain \etal 1997). 

In Figure 2 we show the predicted Type Ia and II(+Ib/c) rest-frame
frequencies as a function of redshift. Expressed in SNu, the Type II rate
is basically proportional to the ratio between the UV and blue galaxy
luminosity densities, and is therefore independent of cosmology. Unlike the SN
frequency per unit volume, which will trace the evolution of the stellar 
birthrate, the  frequency of Type II events per unit blue luminosity
is a monotonic increasing function of redshift, and depends only weakly on the 
assumed star formation history. The Type Ia rates plotted in the figure
assume characteristic ``delay'' timescales after the collapse of the primary 
star to a WD equal to $\tau=0.3, 1$ and 3 Gyr, which virtually encompass 
all relevant possibilities. The SN Ia explosion efficiency was left as an
adjustable parameter to reproduce the observed ratio of SN II to SN Ia
explosion rates in the local universe, ${\rm SNR}_{\rm II}/{\rm SNR}_{\rm
Ia}\approx 3.5$, $5\%<\eta<10\%$ for the adopted models. 

It appears that observational determinations of the SN~Ia rate at $z\sim1$ can 
unambiguously identify the appropriate delay time.  In particular, we 
estimate that measuring the frequency of Type Ia events at both $z\sim0.5$ 
and $z\sim 1$ with an error of 20\% or lower would allow one to determine 
this timescale to within about 30\%.  This kind of observations are by no 
means prohibitive, and these goals could be achieved within a couple of 
years. In fact, ongoing searches for high-$z$ SNe (Perlmutter \etal 1998; 
Garnavich \etal 1998) are currently able to discover and study about a 
dozen new events per observing session in the redshift range 0.4-1.0, and the 
observations are carried out at a rate of about four sessions a year. Since 
determining the frequency of SN~Ia with a 20\% uncertainty requires 
statistics on more than 25 objects per redshft bin, it is clear that, barring 
systematic biases, those rates will soon be know with high accuracy. Also note 
how, relative to the merging scenario, the monolithic collapse model predicts 
Type Ia rates (in SNU) that are, in the $\tau=0.3\,$ Gyr case, a factor of 1.6 
and 4.9 higher at $z=2$ and 4, respectively, with even larger factors found 
in the case of longer delays.  Therefore, once such timescale is calibrated
through the observed ratio SNR$_{\rm Ia}(z=0)/$SNR$_{\rm Ia}(z=1)$, 
one should be able to constrain the star formation 
history of the early universe by comparing the predicted SN~Ia rate at $z>2$
with the observations.

\section{Conclusions}

We have investigated the link between SN statistics and galaxy evolution. 
Using recent determinations of the star formation history of field galaxies
from the present epoch to high-$z$, and a simple model for the evolutionary
history of the binary system leading to a Type Ia event, we have computed the
theoretical Type Ia and Type II SN rates as a function of cosmic time. While
significant uncertainties still remain in these estimates, we believe the 
calculations presented in this {\it Letter} offer a first, realistic glimpse
to the evolution of the cosmic supernova rates with cosmic time. 
Our main results can be summarized as follows.

\begin{itemize}

\item At the present epoch, the predicted Type II(+Ib/c) frequency 
appears to match remarkably well the observed local value. The obvious caveat 
is the well known fact that the SN II rate is a sensitive function 
of the lower mass cutoff of the 
progenitors, $m_l$. Values as low as $m_l=6\,\msun$ (Chiosi, Bertelli, \&
Bressan 1992) or as high as  $m_l=11\,\msun$ (Nomoto 1984) have been proposed 
in the literature: adopting a lower  mass limit of 6 or 11 $\msun$ would
increase or reduce our Type II rates by a factor 1.5, respectively.
Note also that rates obtained from traditional distant (beyond 4 Mpc) sample 
might need to be increased by a factor of 1.5--2 because of severe selection
effects against Type II's fainter than $M_V=-16$ (Woltjer 1997).

\item In the interval $0\lta z\lta 1$, the predicted rate of SN Ia  
is a sensitive function of the characteristic delay timescale between 
the collapse of the primary star to a WD and the SN event. Accurate 
measurements of SN rates in this redshift range will improve 
our understanding of the nature of SN~Ia progenitors and the physics of 
the explosions. Ongoing searches and studies of distant SNe should soon 
provide these rates, allowing a universal calibration of the Type Ia phenomenon.

\item While Type Ia rates at $1\lta z\lta 2$ will offer valuable information 
on the star formation history of the universe at earlier epoch, 
the full picture will only be obtained with statistics on Type Ia and
II SNe at redshifts $2<z<4$ or higher. At these epochs, the  
detection of Type II events must await the
{\it Next Generation Space Telescope} (NGST). A SN II has a typical peak
magnitude $M_B\approx -17$ (e.g. Patat \etal 1994): placed at $z=3$, such an
explosion would give rise to an observed flux of 15 nJy (assuming a
flat cosmology with $q_0=0.5$ and $H_0=50\,h_{50}\kmsmpc$) at 1.8 \micron.
At this wavelength,
the imaging sensitivity of an 8m NGST is 1 nJy ($10^4$ s exposure and
$10\sigma$ detection threshold), while the moderate resolution ($\lambda/\Delta
\lambda=1000$) spectroscopic limit is about 50 times higher ($10^5$ s exposure
per resolution element and $10\sigma$ detection threshold) (Stockman \etal
1998). The several weeks period of peak rest-frame blue luminosity would be
stretched by a factor of $(1+z)$ to few months. Figure 3 shows the cumulative
number of Type II events expected per year per $4'\times 4'$ field. Depending
on the history of star formation at high redshifts, the NGST should detect
between 7 (in the merging model) and 15 (in the monolithic collapse scenario)
Type II SNe per field per year in the interval $2<z<4$. The
possibility of detecting Type II SNe at $z\gta5$ from an early population of
galaxies has been investigated by Miralda-Escud\'e \& Rees (1997). By assuming
these are responsible for the generation of all the metals observed in the
Lyman-$\alpha$ forest at high redshifts, a high baryon density
($\Omega_bh_{50}^2=0.1$), and an average metallicity of $0.01Z_\odot$,
Miralda-Escud\'e \& Rees estimate the NGST should observe about 16 SN II per
field per year with $z\gta 5$. Note, however, that a metallicity smaller by a
factor $\sim 10$ compared to the value adopted by these authors has been
recently derived by Songaila (1997). For comparison, the models discussed in
this {\it Letter} predict between 1 and 10 Type II SNe per field per year with
$z\gta 4$. 

\end{itemize}

\acknowledgments

PM acknowledges support from NASA through ATP grant NAG5-4236. MDV thanks the 
ospitality of the STScI, where part of this work was done. 

\references

Baugh, C.~M., Cole, S., Frenk, C.~S., \& Lacey, C.~G. 1998, \apj, 
in press (astro-ph/9703111)

Branch, D., Livio, M., Yungelson, L.~R., Boffi, F., \& Baron, 
E. 1995, PASP, 107, 1

Bruzual, A.~G., \& Charlot, S. 1998, in preparation

Cappellaro, E., Turatto, M., Tsvetkov, D.~Yu., Bartunov, O.~S.,
Pollas, C., Evans, R., \& Hamuy, M. 1997, A\&A, 322, 431

Chiosi, C., Bertelli, G., \& Bressan, A. 1992, ARA\&A, 30, 235


Connolly, A.~J., Szalay, A.~S., Dickinson, M.~E., SubbaRao, M.~U., 
\& Brunner, R.~J. 1997, \apj, 486, L11

Cowie, L.~L., Songaila, A., Hu, E.~M., \& Cohen, J.~G. 1996,
\aj, 112, 839

Eggen, O.~J., Lynden-Bell, D., \& Sandage, A.~R. 1962, \apj, 136, 748

Ellis, R.~S., Colless, M., Broadhurst, T., Heyl, J., \& Glazebrook,
K. 1996, \mnras, 280, 235

Evans, R., van den Bergh, S., \& McClure, R.D. 1989, \apj, 345, 752

Gardner, J.~P., Sharples, R.~M., Frenk, C.~S., \& Carrasco, B.~E.
1997, \apjl, 480, L99 

Garnavich, P. M., \etal 1998, ApJ, 493, L53

Greggio, L., \& Renzini, A. 1983, \aap, 118, 217


Hansen, L., Jorgensen, H. E., Norgaard-Nielsen, H. U., Ellis, R. S.,
\& Couch, W. J. 1989, \aap, 211, L9

Heyl, J., Colless, M., Ellis, R.~S., \& Broadhurst, T. 1997,
\mnras, 285, 613

Iben, I. Jr., \& Renzini, A. 1983, ARA\&A, 21, 271

Iben, I. Jr., \& Tutukov, A. 1984, ApJS, 54, 535

Kauffmann, G., White, S.~D.~M., \& Guiderdoni, B. 1993, \mnras, 264,
201 


Kim, A.~G., \etal 1997,  \apj, 476, L63  

Lilly, S.~J., Le F{\'e}vre, O., Hammer, F., \& Crampton, D.
1996, \apj, 460, L1

Lilly, S.~J., Tresse, L., Hammer, F., Crampton, D., \& Le F{\'e}vre,
O. 1995, \apj, 455, 108


Madau, P. 1997, in Star Formation Near and Far, ed. S. S. Holt 
\& G. L.  Mundy, (AIP: New York), p. 481

Madau, P., Ferguson, H.~C., Dickinson, M.~E., Giavalisco, M., 
Steidel, C.~C., \& Fruchter, A. 1996, \mnras, 283, 1388

Madau, P., Pozzetti, L., \& Dickinson, M.~E. 1998, \apj, in press
(astro-ph/9708220)



Miralda-Escud\'e, J., \& Rees, M. J. 1997, \apj, 478, L57

Nomoto, K. 1984, \apj, 277, 791

Norgaard-Nielsen, H. U., Hansen, L., Jorgensen, H. E., Arag{\'o}n-Salamanca, A.,
\& Ellis, R. S. 1989, Nature, 339, 523

Pain, R., \etal 1997, ApJ, 473, 356


Patat, R., Barbon, R., Cappellaro, E. Turatto, M. 1994, A\&A, 282, 731


Perlmutter, S., \etal 1998, Nature, 391, 51

Pettini, M., Smith, L.~J., King, D.~L., \& Hunstead, R.~W. 1997, \apj, 486,
665

Pettini, M., Steidel, C.~C., Dickinson, M., Kellogg, M., 
Giavalisco, M., \& Adelberger, K.~L. 1998, in The Ultraviolet Universe at Low 
and High Redshift, ed. W. Waller, (Woodbury: AIP Press), in press 
(astro-ph/9707200)

Renzini, A. 1997, \apj, 488, 35

Ruiz-Lapuente, P., Canal, R., \& Burkert, A. 1997, in 
Thermonuclear Supernovae, ed. P. Ruiz-Lapuente, R. Canal, \& J. Isern 
(Dordrecht: Kluwer), p. 205

Salpeter, E.~E. 1955, \apj, 121, 161

Schechter, P.~L., \& Dressler, A. 1987, \aj, 94, 56

Songaila, A. 1997, \apj, 490, L1

Steidel, C.~C., Adelberger, K.~L., Dickinson, M.~E., Giavalisco, M., 
Pettini, M., \& Kellogg, M. 1998, \apj, 492, 428

Steidel, C.~C., Giavalisco, M., Pettini, M., Dickinson, M.~E., 
\& Adelberger, K.~L. 1996, \apj, 462, L17

Stockman, H. S., Stiavelli, M., Im, M., \& Mather, J. C. 1998, in 
Science with the Next Generation Space Telescope, ed. E. Smith \&
A. Koratkar (ASP Conf. Ser.), in press

Tammann, G.~A., L\"offler, W., \& Schr\"oder, A. 1994, ApJS, 92, 487

Tinsley, B.~M., \& Gunn, J.~E. 1976, \apj, 203, 52

Treyer, M.~A., Ellis, R.~S., Milliard, B., \& Donas, J. 1998,
in The Ultraviolet Universe at Low and High Redshift, ed. W. Waller, 
(Woodbury: AIP Press), in press (astro-ph/9706223)

van den Bergh, S., \& Tammann, G.~A. 1991, ARA\&A, 29, 363

Weidemann, V. 1987, \aap, 188, 74

Wheeler, J. C., \& Swartz, D. A. 1993, \ssr, 66, 425

Whelan, J., \& Iben, I. Jr. 1973, \apj, 186, 1007

White, S.~D.~M., \& Frenk, C.~S. 1991, \apj, 379, 25

Williams, R.~E., \etal 1996, \aj, 112, 1335

Woltjer, L. 1997, A\&A, 328, L29 

Yungelson, L., \& Livio, M. 1998, \apj, in press (astro-ph/9711201)

\clearpage 
\begin{deluxetable}{lccccccccr}
\tabcolsep 0.0pt
\tablewidth{6.0in}
\tablenum{1}
\tablecaption{Present-epoch SN Rates\tablenotemark{a} \tablenotemark{b} \label{lumden}}
\tablehead{
\colhead{Hubble Type} & \colhead{Ia} & \colhead{Ib/c} & \colhead{II} 
}
\startdata
E--S0 & 0.07, 0.25, 0.08 & \nodata  & \nodata
\nl
S0a--Sa & 0.11, 0.12, 0.15 & 0.07, 0.01, \nodata & 0.07, 0.04, \nodata 
\nl
Sab--Sb    & 0.08, 0.12, 0.08 & 0.04, 0.07, 0.15   & 0.24, 0.34, 0.45 
\nl
Sbc--Sd    & 0.11, 0.12, 0.05 & 0.07, 0.19, 0.18   & 0.39, 0.98, 0.58
\nl
\enddata
\tablenotetext{a}{Values listed are in units of $h_{50}^2$ SNu (1 SNu= 1 SN per
century per $10^{10} L_{B\odot}$).}
\tablenotetext{b}{The first column in each entry shows the value determined by
Cappellaro \etal (1997), second column by Tammann \etal (1994), third column by
Evans \etal (1989).} 
\end{deluxetable}
\clearpage

\begin{figure}
\plotone{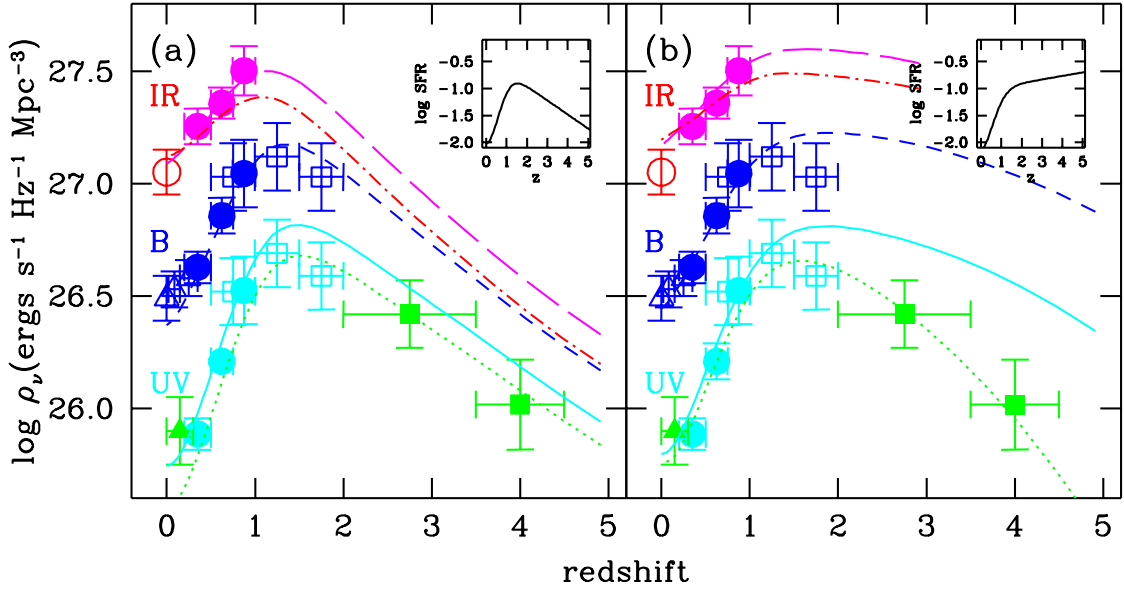}
\vspace{-3.0cm}
\caption{Evolution of the observed comoving luminosity density at rest-frame 
wavelengths of 0.15 ({\it dotted line}), 0.28 ({\it solid line}), 0.44 
({\it short-dashed line}), 1.0 ({\it long-dashed line}), and 2.2 ({\it 
dot-dashed line}) \micron. The data points with error bars are taken from 
Lilly \etal (1996) ({\it filled dots}), Connolly \etal (1997) ({\it empty 
squares}), Madau \etal (1996) and Madau (1997) ({\it filled squares}), Ellis 
\etal (1996) ({\it empty triangles}), Treyer \etal (1998) ({\it filled 
triangle}), and Gardner \etal (1997) ({\it empty dot}). A flat cosmology with
$q_0=0.5$ and $H_0=50\,h_{50}\kmsmpc$ was adopted. The inset in the upper-right
corner of the plot show the SFR density ($\sfrd$) vs. redshift which was used 
as input to the population synthesis code. ({\it a}) The model 
assumes a Salpeter IMF, SMC-type dust in a foreground screen (with the 
scattering term removed), and a universal $E_{\rm B-V}=0.1$. 
({\it b}) Test case with a much larger star formation density at high redshift
than indicated by the HDF dropout analysis. The model -- designed to mimick a
``monolithic collapse'' scenario -- assumes a Salpeter IMF and a dust opacity
which increases rapidly with redshift, $E_{\rm B-V}=0.011(1+z)^{2.2}$. 
\label{fig1}}
\end{figure}

\begin{figure}
\plotone{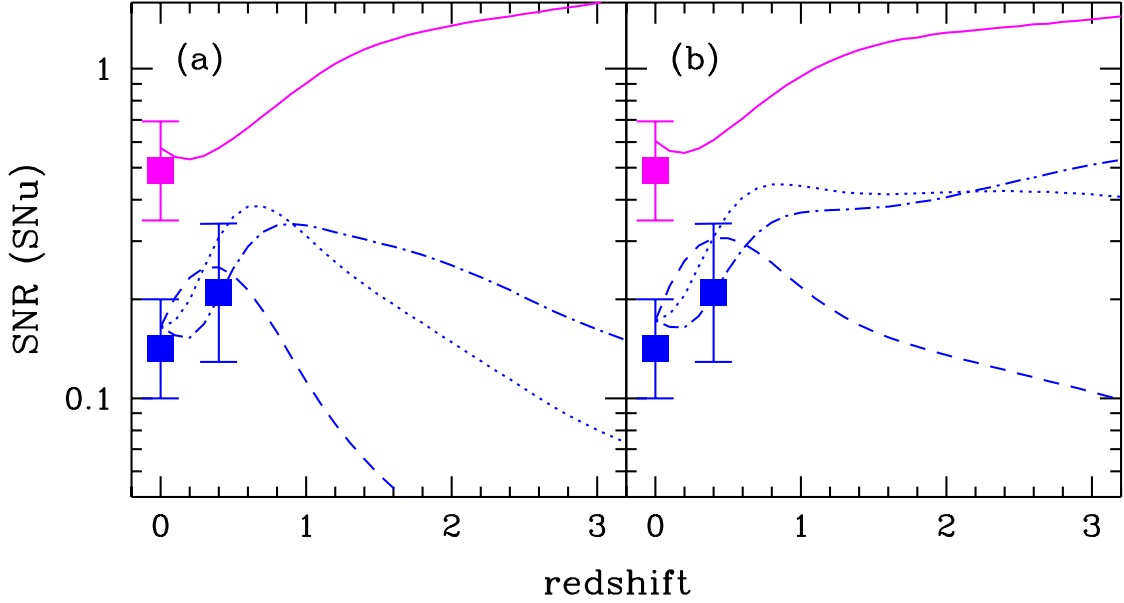}
\vspace{-3.0cm}
\caption{Predicted Type Ia and II(+Ib/c) rest-frame frequencies as a function 
of redshift. The rates are normalized to the {\it emitted} blue luminosity 
density. {\it Solid line}: SN II rate.	{\it Dashed-dotted line}: SN Ia 
rate with $\tau=0.3$ Gyr. {\it Dotted line}: SN Ia rate with $\tau=1$ Gyr. 
{\it Dashed line}: SN Ia rate with $\tau=3$ Gyr. The data points with error
bars have been derived from the measurements of Cappellaro \etal (1997),
Tammann \etal (1994), Evans \etal (1989), and Pain \etal (1997). ({\it a})
Model predictions for the ``merging'' scenario of Figure 1{\it a}.  ({\it b})
Same for the ``monolithic collapse'' scenario of Figure 1{\it b}. 
\label{fig2}}
\end{figure}

\begin{figure}
\plotone{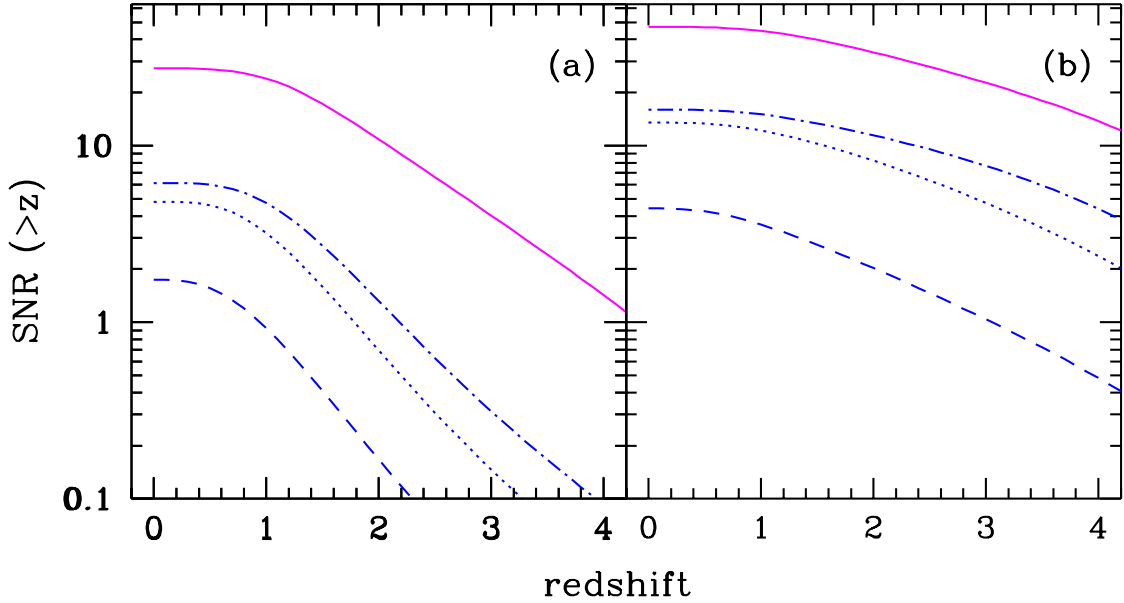}
\vspace{-3.0cm}
\caption{Predicted cumulative number of Type Ia and II(+Ib/c) SNe above a
given redshift $z$ in a $4'\times 4'$ field. {\it Solid line}: Type II's.
{\it Dashed-dotted line}: Type Ia's with $\tau=0.3$ Gyr. {\it Dotted 
line}: Type Ia's with $\tau=1$ Gyr. {\it Dashed line}: Type Ia's with 
$\tau=3$ Gyr. The effect of dust extinction on the detectability of SNe is 
negligible in these models. ({\it a}) Model predictions for the ``merging''
scenario of Figure 1{\it a}. ({\it b}) Same for the ``monolithic collapse''
scenario of Figure 1{\it b}. 
\label{fig3}}
\end{figure}

\end{document}